\documentclass[prd,twocolumn,superscriptadress,showpacs,showkeys]{revtex4}

\usepackage{graphicx}
\usepackage{amssymb}
\usepackage{amsmath}
\usepackage{fancyhdr}
\usepackage{esint}

\newcommand{\PTm}{\ensuremath{\mathcal{PT}}}

\newcommand{\oper}[1]{\ensuremath{\mathsf{#1}}}

\newcommand{\be}{\begin{equation}}
\newcommand{\en}{\end{equation}}
\newcommand{\bea}{\begin{eqnarray}}
\newcommand{\ena}{\end{eqnarray}}
\newcommand{\ee}{\ensuremath{\mathrm{e}}}
\newcommand{\dd}{\ensuremath{\mathrm{d}}}
\newcommand{\ii}{\ensuremath{\mathrm{i}}}

\newcommand{\abs}[1]{\ensuremath{\left|#1\right|}}
\newcommand{\odk}[1]{(\ref{#1})}

\newcommand{\ket}[1]{\ensuremath{|#1\rangle}}
\newcommand{\bra}[1]{\ensuremath{\langle#1|}}

\newcommand{\ssd}[2]{\ensuremath{\langle#1|#2\rangle}}

\newcommand{\ve}{\varepsilon}
\newcommand{\vf}{\varphi}
\newcommand{\vt}{\vartheta}
\newcommand{\odkf}[1]{Fig.~\ref{#1}}
\newcommand{\ks}[1]{#1^*}

\newcommand{\re}{\mathop{\mathrm{Re}}}
\newcommand{\im}{\mathop{\mathrm{Im}}}

\newcommand{\bes}{\begin{subequations}\bea}
\newcommand{\ens}{\ena\end{subequations}}

\begin{document}

\title{Adiabatic time-dependent metrics in \PTm-symmetric quantum theories}

\author{Hynek B\'ila}
\email{bila@ujf.cas.cz}
\affiliation{Department of Theoretical Physics, Nuclear Physics Institute, \v Re\v z
(Prague), Czech Republic}
\affiliation{Faculty of Mathematics and Physics, Charles University, Prague, Czech
Republic}

\begin{abstract}
We introduce an approach to scattering problems in theories with non-Hermitian Hamiltonian, usually known as \PTm-symmetric quantum theories, by means of the adiabatic switching of the interaction. The modifications of usual methods needed to employ time-dependent metrics are described. We argue that an analogue of the adiabatic theorem hold for time dependent metrics and that its validity forms a necessary condition for consistency of the procedure. Two toy models are presented for sake of illustration.
\end{abstract}


\maketitle

\section{Introduction}

Quantum theories whose evolution was driven by a non-Hermitian Hamiltonian have a long history emerging in the early times of quantum physics (for the earliest accounts see \cite{pauli1943} where the indefinite metric quantum theories are discussed in detail; modern approaches are based on treatment of \cite{SGH}). They became a prominent topic of discussions in last decade after publication of the key article \cite{PTSQM} mostly under the name \PTm-symmetric quantum theories. A vital point of modern interpretations of such theories is always the redefinition of the scalar product which, assuming real spectrum of the Hamiltonian, enables to re-establish the consistent probabilistic interpretation of the theory. The standard machinery goes as follows: At first, one takes the theory's non-Hermitian Hamiltonian $H$ and check whether it has real spectrum. If so, one tries to find an operator $\Theta$, called a metric, which satisfies
\be
H^\dag\Theta=\Theta H.\label{quasi}
\en
The metric has to be positive, bounded, invertible with bounded inverse, Hermitian operator, and must map the range of $H$ into the domain of definition of $H^\dag$. (For Hamiltonians with complex spectrum such metric does not exist.) Having found $\Theta$ one may define new scalar product
\be
\ssd{\Psi}{\Phi}_\Theta=\ssd{\Psi}{\Theta\Phi}
\en
which should be used for calculating transition amplitudes. Equivalently, if one is able to decompose the metric into
\be
\Theta=\Omega^\dag\Omega,
\en
one can define a new, Hermitian Hamiltonian by a similarity transformation as
\be
h=\Omega H\Omega^{-1}.
\en

If \odk{quasi} holds for some operator $A$ in place of $H$, we say that $A$ is quasi-Hermitian with respect to $\Theta$. All observables of the theory have to be quasi-Hermitian with respect to the same metric. For any quasi-Hermitian observable we can of course define its Hermitian counterpart $a=\Omega A\Omega^{-1}$. Although the Hermitian representation using $h$ as a Hamiltonian is assumed to exist for any model, for practical purposes it is often preferable to make calculations in the original basis using the metric, since the $h$ (and eventually other observables) tend to be rather complicated and inconvenient to deal with.

It is not difficult to see that the probabilities are conserved under described setting, for we have
\be
\frac{\dd}{\dd t}\ssd{\Psi}{\Theta\Phi}=\ii\ssd{H\Psi}{\Theta\Phi}-\ii\ssd{\Psi}{\Theta H\Phi}
\label{tderm}
\en
which vanishes due to quasi-Hermiticity condition. The quasi-Hermiticity condition can be rewritten in terms of the evolution operator
\be
U(t',t)=\ee^{-\ii H(t'-t)}.
\en
Instead of standard $U^\dag U=I$ one has now
\be
U^{\dag}\Theta U=\Theta.
\en

It may be noted that the equation \odk{quasi} does not define the metric uniquely. To specify a unique metric, one has to employ the quasi-Hermiticity condition for other observables; if the set of observables is irreducible there can be only one metric which fulfills the given requirements. The situation can become more peculiar when the Hamiltonian is time-dependent. In such a case there is a continuous set (depending on $t$) of equations
\be
H(t)^\dag\Theta=\Theta H(t)
\en
that have to be satisfied. The set $\{H(t)\}$ can well be irreducible and specify the metric uniquely, but it is entirely possible for the metric to be over-specified and thus non-existent. This leads to the concept of quasi-stationarity: the Hamiltonian is said to be quasi-stationary if there exists a single metric with respect to which it is quasi-Hermitian at any $t$. Only quasi-stationary Hamiltonians are accessible for the standard physical interpretations. Note that one cannot simply assume a time-dependent metric since it would lead to an additional term in \odk{tderm} proportional to $\dd\Theta/\dd t$. We will discuss how to overcome these problems in section \ref{tdmsect}.

\section{Scattering in non-Hermitian theories}

The number of \PTm-symmetric models investigated in recent literature is vast. Most of the research has been concerned mainly with spectral properties of the systems and especially the presence of bound states. Whilst the treatment of bound states has been standardised and became a routine task, the discussion of scattering has faced several conceptual difficulties. The problems are closely related to the metric redefinition, of course. To see the core of the complications, let us consider two special cases:

At first we look more closely to quantum-mechanical scattering with short-range potential. The standard approach to the problem uses $x$-representation, solving the Schr\"odinger equation with plain-wave-like boundary conditions for specification of the incoming wave at large $\abs{x}$ in a chosen direction; the scattering amplitudes are taken as the asymptotics of the solution again at large $\abs{x}$. This is reasonable since in asymptotic regions (out of range of the interaction) the solutions from the continuous spectrum solve the free-particle Schr\"odinger equation and thus can be interpreted as free particles coming into (or out from) the interaction region.

Such approach can hardly be justified with non-Hermitian interaction. The reason is that $x$ in the wave function $\psi(x)$ has no longer the interpretation of an eigenvalue of the position operator. This is mainly due to the scalar product redefinition: the operator $X$ acting as $[X\psi](x)=x\psi(x)$ is generically not $\Theta$-quasi-Hermitian and therefore not acceptable as an observable. In order to find the real asymptotic states one has to construct physical position operator first. This can prove difficult and even impossible in some situations, namely within relativistic quantum mechanics \footnote{See e.g. \cite{AliHSS} for non-Hermitian description of free Klein-Gordon equation, which follows the course laid out in \cite{FV}. The reasons for non-existence of the position operator are not connected with indefiniteness of the metric and consequent problems reappear in non-Hermitian formulation.}.

\newcommand{\Hi}{H_\mathrm{I}}
When the use of $x$-representation is inconvenient (which includes most of the field theory, relativistic and many-particle physics), it became customary to use different approach, which is based on decomposition of the Hamiltonian into free part $H_0$ and interaction part $\Hi$. The scattering amplitudes are the matrix elements of the S-matrix between the free Hamiltonian eigenstates; S-matrix is given by
\be
S=\lim_{t_\pm\rightarrow\pm\infty}U_0(0,t_+)U(t_+,t_-)U_0(t_-,0)
\label{smlim}
\en
where the indices differentiate between free and full evolution operators in the same way as for the Hamiltonian. The sense of the convergence in \odk{smlim} has to be clarified. For physical reasons we can expect the limit to exist for particular matrix elements. In many-particle systems these are the matrix elements between particle states whose momentum-representation wave functions have disjoint supports at $t=0$; this is because for large $\abs{t}$ the free evolution operator removes them far apart and hence out from the range of interaction; after that they evolve exactly the same way under $U_0$ as under $U$ and the contributions of large $\abs{t}$ on the right-hand side of \odk{smlim} cancel.

With a non-Hermitian $\Hi$ the situation is more complicated due to the necessity of scalar product redefinition. The main problem is usually the fact that $H$ and $H_0$, although both quasi-Hermitian, are not so with respect to the same metric. This represents a serious obstacle because \odk{smlim}, however formally treated, would lead to an S-matrix unitary with respect neither to $\Theta_0$ nor to $\Theta$ ($\Theta_0$ is the free metric, i.e. calculated for the free Hamiltonian, and often it is simply unit operator).

A useful trick which helps to get better meaning to \odk{smlim} is the introduction of adiabatic switching of the interaction. This consists of introduction of an artificial time-dependent damping factor to the interaction Hamiltonian. The exact form of this factor is not important as long as its derivative is small and the factor is equal to 1 at $t=0$ and tends to zero at infinities. For concreteness one can take
\be
H_\ve(t)=H_0+\ee^{-\ve\abs{t}}\Hi
\label{heps}
\en
with $\ve$ small positive. Hence for large $\abs{t}$ the full and free Hamiltonians are essentialy the same and the limit \odk{smlim} may exist on larger domain \footnote{Without the damping factor the limit exists (in strong operator sense) only on the domain of scattering states, or put more rigorously, as $\mathop{\textrm{s-lim}}_{t\rightarrow\infty}U(t)^\dag U_0(t)P$ where $P$ is spectral projector on the absolutely continuous spectrum of $H_0$.}. The change presumably does not affect the scattering amplitudes between particles that are outside the range of interaction when $\ee^{-\ve\abs{t}}$ differs significantly from one. The original scattering problem can be recovered by putting $\ve\rightarrow 0$ \footnote{In perturbative treatment of field theories the most important part is that the free Hamiltonian ground state (vacuum) is mapped to the full Hamiltonian vacuum, whose validity was demonstrated in appendix to \cite{GML} (Gell-Mann and Low theorem).}. 

Analogous approach to the problem of scattering for non-Hermitian $H$ would face all the problems of time-dependent Hamiltonians described in the introduction. Since in general there is little chance to find common metric for $H$ and $H_0$ (and thus for all $H_\ve(t)$ producing the quasi-stationary scenario) we have to develop physical interpretation for systems with time-dependent metrics.

\section{Time-dependent metrics}
\label{tdmsect}

It turns out that in order to evade the demand of quasi-stationarity one cannot maintain $H$ both as an observable and as the time evolution generator. In the latter case one finds the metric by solving \odk{quasi} for each $t$ separately and then postulates the Schr\"odinger equation in the Hermitian representation
\be
\ii\frac{\dd\psi(t)}{\dd t}=h(t)\psi(t)
\label{hermschr}
\en
(we keep the distinction between small and capital letters, which distinguishes the original and the Hermitian representation, for the wave functions too, i.e. $\psi=\Omega\Psi$). One must now sacrifice the validity of
\be
\ii\frac{\dd\Psi(t)}{\dd t}=H(t)\Psi(t)
\label{nhermschr}
\en
since for time-dependent $\Theta$ the equations \odk{hermschr} and \odk{nhermschr} are not equivalent. Instead of $H$, the time evolution in the non-Hermitian representation is now driven by
\be
H_\mathrm{gen}=H-\ii\Omega^{-1}\dot{\Omega}
\label{hgen}
\en 
where dot denotes the time derivative. This resolution was suggested in \cite{ZnojilCZ}.

Alternatively, one can maintain \odk{nhermschr} and postulate the probability conservation
\be
\frac{\dd}{\dd t}\ssd{\Psi}{\Theta\Phi}=0
\label{unita}
\en
for all $\Psi$ and $\Phi$. Combining \odk{nhermschr} and \odk{unita} one gets
\be
\dot{\Theta}=\ii(\Theta H-H^\dag\Theta).\label{tgr}
\en
Hence, the algebraic equation \odk{quasi} is replaced by differential equation \odk{tgr}. The condition \odk{tgr} is needed only for the Hamiltonian, the observables must still satisfy ordinary quasi-Hermiticity condition. Consequently, if there is not a static (time independent) solution to \odk{tgr}, $H$ is no more an observable. However, we can construct an ``observable Hamiltonian''
\be
H_\mathrm{obs}=H+\ii\Theta^{-1}\dot{\Theta}.
\label{hobs}
\en
If necessary, an equivalent Hermitian representation can be obtained. It has no sense to transform $H$, rather one shall apply the similarity transformation to the evolution operator $U$ which yields
\be
u(t,t')=\Omega(t)U(t,t')\Omega(t')^{-1}.
\en
Since we have 
\be
h=\ii\frac{\dd u(t,t')}{\dd t}u(t,t')^{-1}
\en
we can easily realise that
\be
h=\Omega H_\mathrm{obs}\Omega^{-1}.\label{hercp}
\en

The transition probability amplitudes are in this framework computed as $\Theta(t)$-induced scalar product of concerned Schr\"odinger-picture state vectors taken in the same time instant as the metric. For instance the probability of finding state $\Phi$ at time $t$ in a system prepared in state $\Psi$ at time $t'$ is
\be
P=\abs{\bra{\Phi}\Theta(t)U(t,t')\ket{\Psi}}^2=\abs{\bra{\Phi}U(t',t)^\dag\Theta(t')\ket{\Psi}}^2.
\label{probab}
\en

Let us note that the non-observable nature of $H$ is in fact not a grave problem. When the Hamiltonian is time-dependent, the energy is no more conserved and cannot be directly measured even one we deals with completely conventional Hermitian system. The energy such a system has well-defined approximative meaning only if the time dependence is weak. In that case one can as well use $H_\mathrm{obs}$ provided $\dot{\Theta}$ is small. That such slowly changing $\Theta$ exists is not a priori clear and we will return to this point in later discussion.

The former approach with the observable $H$ is suitable for situations where one intends to use the non-Hermitian Hamiltonian as an auxiliary representation of a time-dependent Hermitian $h$ which is given as the primary operator. If the theory is formulated with $H$ as a generator of time evolution (as it clearly is in case of the adiabatic interaction switching) one must accept the latter resolution, however it is more complicated. Compared treatments are not so different as it could appear on the first sight: The similarity of formul\ae{} \odk{hgen} and \odk{hobs}, as well as the validity of \odk{hercp} suggest that by identifying $H$ (of the former approach) with $H_\mathrm{obs}$ and $H_\mathrm{gen}$ with $H$ (of the latter approach) an isomorphism between these two methods can be established; it only matters where one starts and what has to be calculated. For reasons given above we will exclusively treat $H$ as a time evolution generator in the following.

\subsection*{Calculation of the metric}

Obviously the calculation of the metric based on \odk{tgr} is complicated and has to be done numerically or approximatively in majority of conceivable models. The apparent similarity between \odk{tgr} and the Schr\"odinger equation for the evolution operator or the Heisenberg equations for observables suggests we can use similar techniques, e.g. iteration of the equation rewritten as
\be
\Theta(t)=\Theta(0)+\int_0^t \Theta(t')H(t')-H(t')^{\dag}\Theta(t')\dd t'.\label{iter}
\en
Contrary to the Schr\"odinger equation the process is further complicated by the ambiguity in setting $\Theta(0)$, which can be resolved only with more detailed knowledge of the specific model. However, if $\Theta(0)$ is chosen Hermitian, so is $\Theta(t)$ due to the right-hand side of \odk{tgr}. Of course there is a possibility not to calculate $\Theta$ directly and find the evolution operator instead. The metric is then, as suggested by \odk{probab},
\be
\Theta(t)=U(t',t)^\dag\Theta(t')U(t',t).
\en

\newcommand{\Ts}{\Theta^\mathrm{S}}
Although it is uneasy to give details about the metric in general case, hopefully we can say more about systems with slowly changing $H(t)$ which are the main topic of interest in this paper. Even more specific case is that of time-independent Hamiltonian: one may ask about the solutions of \odk{tgr} when $H$ does not depend on time. If $H$ is quasi-Hermitian, which we will always suppose even if we can consistently solve \odk{tgr} for broader class of Hamiltonians, then there is always, by definition of quasi-Hermiticity, a static solution which we will denote $\Ts$. But in addition there are other, non-static solutions. In finite-dimensional Hilbert space, there is always a basis in which their matrix elements are periodic. Indeed, if we write the metric as
\be
\Theta=\vt_{mn}\ket{\Psi^m}\bra{\Psi^n}
\en
where $\Psi^n$ is the $n$-th eigenvector of $H^\dag$ and $\vt_{mn}^{}=\vt_{nm}^*$, \odk{tgr} gives
\be
\dot{\vt}_{mn}=\ii(E_m-E_n^*)\vt_{mn}\label{organon}
\en 
which leads to constant diagonal elements and periodic off-diagonal elements. We have used the biorthonormality relations between $\Psi^n$ and the eigenvectors of $H$ which we denote $\Psi_n$:
\be
\ssd{\Psi^m}{\Psi_n}=\delta_{mn}.\label{biort}
\en
The whole metric is periodic if there is a common multiple of all energy differences. Note that if $H$ loses its real spectrum $\vt_{mn}$ will exponentially rise or decay (diagonal elements included) and the periodicity will be lost. Under specific conditions the argument can be extended to infinite-dimensional diagonalisable Hamiltonians.

For constant Hamiltonian one can formally solve the equation \odk{iter}. In special case with $\Theta(0)=I$ one gets
\bea
\Theta(t)&=&I+\ii(H-H^\dag)t-(H^2-2H^\dag H+{H^\dag}^2)\frac{t^2}{2}+\dots\nonumber\\
&=&\ :\!\exp\ii(H-H^\dag)t\!:
\label{normex}
\ena
the colon in the last term standing for normal ordering of the exponential's Taylor series, which means all $H^\dag$ standing left of any $H$ in each monomial. With different starting metric one has to insert $\Theta(0)$ in each term between the rightmost $H^\dag$ and the leftmost $H$. Taking truncated series \odk{normex} makes a good approximation if, roughly speaking, $(H-H^\dag)t$ is small.

\subsection*{Time-dependent metrics in scattering problems}

In order to make sense of scattering in theory with adiabatically switched non-Hermitian interaction we need to modify the definition of S-matrix in \odk{smlim} to reflect the presence of the metric. It will be convenient to separate the S-matrix into product of the M\o ller operators
\bes
\Omega^{(+)}&=&\lim_{t\rightarrow\infty}U_0(0,t)U(t,0)\,,\\
\Omega^{(-)}&=&\lim_{t\rightarrow-\infty}U(0,t)U_0(t,0)
\ens
and introduce the customary notation of in- and out-states as
\bes
\Psi^\mathrm{in}&=&\Omega^{(-)}\Psi, \\
\Psi^\mathrm{out}&=&\Omega^{(+)}\Psi. \label{moler}
\ens
Then, the transition amplitudes are given as $\ssd{\Psi_\mathrm{f}^\mathrm{out}}{\Psi_\mathrm{i}^\mathrm{in}}$ and may be easily generalised for use in non-Hermitian theories to
\be
S_\mathrm{fi}=\bra{\Psi_\mathrm{f}^\mathrm{out}}\Theta\ket{\Psi_\mathrm{i}^\mathrm{in}}\,,
\label{sfidef}
\en
subscripts $_\mathrm{f}$ and $_\mathrm{i}$ denoting final and initial state as usually.

A convenient interpretation of in-states (out-states analogously) is that, under limit $t_-\rightarrow-\infty$ they are states evolved first from 0 to $t_-$ under free Hamiltonian (which is motivated by need to get rid of the oscillating phase in $U(t_-,0)$ in the limit) and then back to $t=0$ under full Hamiltonian. Applying the logic of adiabatic switching we shall set the metric at $t=-\infty$ to conform the free Hamiltonian and let it evolve using \odk{tgr}. Hence $\Theta$ in \odk{sfidef} is not an arbitrary metric; on the contrary it is uniquely specified as the metric which adiabatically evolves from the free metric $\Theta_0$ (which we will often assume to be equal to $I$):
\be
\Theta=\lim_{\ve\rightarrow 0} \lim_{t\rightarrow-\infty} U_\ve(t,0)^\dag\Theta_0 U_\ve(t,0),
\label{adiabatheta}
\en
$\ve$ having the same meaning as in \odk{heps}. Note that $U_\ve$ is not unitary and so the right-hand side does not simply evaluate to $I$ even in case of $\Theta_0=I$. 

To make the approach consistent, we have to show that the limit in \odk{adiabatheta} exists and is unique. In particular this implies the following requests:
\begin{enumerate}
	\item When transition between two time-independent Hamiltonians happens adiabatically, a static metric must evolve into another static metric. E.g. changing $H_0$ to $H_1$ during some interval $(0,T)$ with $H_T(t)=\frac{(T-t)}{T}H_0+\frac{t}{T}H_1$ in the meantime, the metric which started as static for $H_0$ must evolve into static metric for $H_1$ when $T\rightarrow\infty$ (the last condition replaces $\ve\rightarrow0$ here). 
	\item The resulting metric should be independent on the precise manner of the switching, i.e. all different damping functions should lead to identical results in the adiabatic limit.
\end{enumerate}

We have seen that for time-independent quasi-Hermitian diagonalisable Hamiltonian the solution to \odk{tgr} is bounded for all $t$ and it oscillates around the static solution $\Ts$. In the (non-orthogonal) eigenbasis of $H^\dag$ the static solutions are diagonal and the non-diagonal elements are sinusoidal. The static solutions are therefore not unstable, i.e. if $\Theta(0)$ is close to any of the static solutions it remains close for all $t$. This is, of course, beneficial: otherwise the smallest deviation from the static metric, which necessarily occurs for finitely slow transition, would lead to utter collapse of the staticity of the metric at later time, and to cure that in the adiabatic limit (i.e. infinitely slow transition) could be impossible.

Our demands resemble somewhat the adiabatic theorem of ordinary quantum mechanics. Recall that the adiabatic theorem says that the eigenvector of $H_0$ evolves into eigenvector of $H_1$ if the transition is adiabatic. Standard proofs use the criterion of adiabaticity which, roughly speaking, says that $\dd{H}/\dd t\ll\delta$ during the whole transition, where $\delta$ is a spectral gap separating the respective eigenvector from the rest of the spectrum. We can translate the theorem to suit our purposes, replacing ``eigenvector'' with ``static metric'', albeit there is an important difference -- the absence of gaps in the set of static metrics.

Although we cannot generalise the proof of adiabatic theorem directly, we can indeed use it to establish the adiabaticity of the metric, even only for Hamiltonians with discrete non-degenerate spectrum. Moreover we will assume that for both $H_0$ and $H$ exists some
\be
\delta=\min_{m,n}\abs{E_n-E_m}
\en
and similar minimum exists for the transitional Hamiltonian $H_\ve$ for all $t$. Then the gap condition holds for sufficiently slow transition and in the adiabatic limit we have
\be
\lim_{\ve\rightarrow0}U_\ve(-\infty,0)\Psi_n=\Phi_n,\label{adiabv}
\en
for all $\Phi_n$ which are eigenvectors of $H_0$ while $\Psi_n$ are eigenvectors of $H$. Thus due to \odk{biort} we can write
\be
U_\ve(-\infty,0)=\sum_n \ket{\Phi_n}\bra{\Psi^n}+O(\ve).
\en
A static metric for $H_0$ can be written as
\be
\Theta_0=\sum_n \vt_n \ket{\Phi^n}\bra{\Phi^n}\label{vts}
\en
(when $H_0$ is Hermitian possibly $\Phi^n=\Phi_n$ and with $\vt_n=1$ the unit metric is recovered). Combining \odk{vts} with \odk{adiabatheta} using \odk{biort} yields
\be
\Theta=\sum_n \vt_n \ket{\Psi^n}\bra{\Psi^n}+O(\ve)\label{vtsu},
\en
i.e. for $\ve\rightarrow0$ a metric which is static because diagonal in the eigenbasis of $H^\dag$. Note that \odk{vtsu} does not specify the metric uniquely since the biorthonormal eigenbasis can have several distinct normalisations and the normalisation is not specified in \odk{adiabv}.

\section{Toy models}

\subsection*{Two-level system}

To illustrate the behaviour of time-dependent metrics we present two different toy models. The first one is a two-level system whose Hamiltonian will be parametrised by Pauli matrices:
\bes
H & = & h_0 + h_i\sigma_i, \\
\Theta & = & \vt_0 + \vt_i\sigma_i,
\ens
with $i\in\{1..3\}$. Equation \odk{tgr} then reads
\bea
\dot\vt_0+\dot\vt_i\sigma_i&=&
2\ii[(\vt_0+\vt_i\sigma_i)\im h_0+(\vt_i+\vt_0\sigma_i)\im h_i]\nonumber\\
&&+2\sigma_i\ve_{ijk}\vt_j\re h_k,
\ena
$\ve$ being the Levi-Civita symbol. After putting 
\bes
2\re h_\mu & = & v_\mu\\ 
2\im h_\mu & = & w_\mu
\ens
with $v,w \in \mathbb{R}$ and $\mu \in \{0..3\}$ we get the following equations
\bes
-\dot\vt_0 & = & \vt_\mu w_\mu \\
-\dot\vt_i & = & \vt_0 w_i+\vt_i w_0+\ve_{ijk}\vt_j v_k.
\ens

Absolutely simplest is the case of Hermitian, time-independent $H$, i.e. $w_\mu=0$. The equations then reduce to
\bes
\dot\vt_0&=&0\\
\dot{\vec\vt}&=&\vec v\times\vec\vt.
\ens
Here the 0th component decouples from the rest and can be adjusted to an arbitrary constant. The vector $\vec\vt$ precesses with constant velocity around the direction specified by the vector $\vec v$. If $\vec\vt(0)$ and $\vec v$ are collinear, then the solution is constant -- in such case $H$ and $\Theta$ commute. If not, the frequency of precession is proportional to $\vec v$, in other words, to the difference between the energies, as it was expected from \odk{organon}, and it does not depend on the initial condition.

The case $w_\mu\neq0$ (i.e. non-Hermitian, but still constant $H$) is analogous. Either it can be transformed by a similarity transformation to the previous case, or managed directly: First we find the static 
\newcommand{\stt}{^\mathrm{S}}
solution $\vt\stt$, if possible. The static solution exists if $\oper{H}$ is pseudo-Hermitian, i.e. if
\bes
w_0&=&0,\\
\vec v\cdot \vec w&=&0.
\ens 
Then,
\be
\vec\vt\stt=-\frac{\vt_0\stt}{\vec v^2}[\vec v\times\vec w]+\alpha \vec v,
\en
where $\vt\stt_0$ and $\alpha$ can be chosen arbitrarily. General solution is easily obtained by setting up the 
Ansatz
\be
\vec\vt=\alpha \vec v+\beta\vec w+\gamma[\vec v\times\vec w].{}
\en
It follows that $\alpha=\mathrm{const.}$ and, up to a time shift,
\be
\beta\propto\sin\sqrt{v^2-w^2}\,t.
\en
The remaining unknowns we get from $\dot\gamma=\beta$ and $\dot\vt_0=-w^2\beta$. This is again what we have expected: if $v^2>w^2$ the energies of $H$ are complex and consequently the periodicity of the metric is replaced by exponential growth.

When the time dependence is enabled we again see the expected behaviour. When the Hamiltonian changes rapidly, the metric is disrupted from its static state and starts to oscillate, whereas under slow transition the oscillations are suppressed. An illustration is given if \odkf{obrasek}.

\begin{figure*}
\begin{center}
\includegraphics[width=0.45\textwidth]{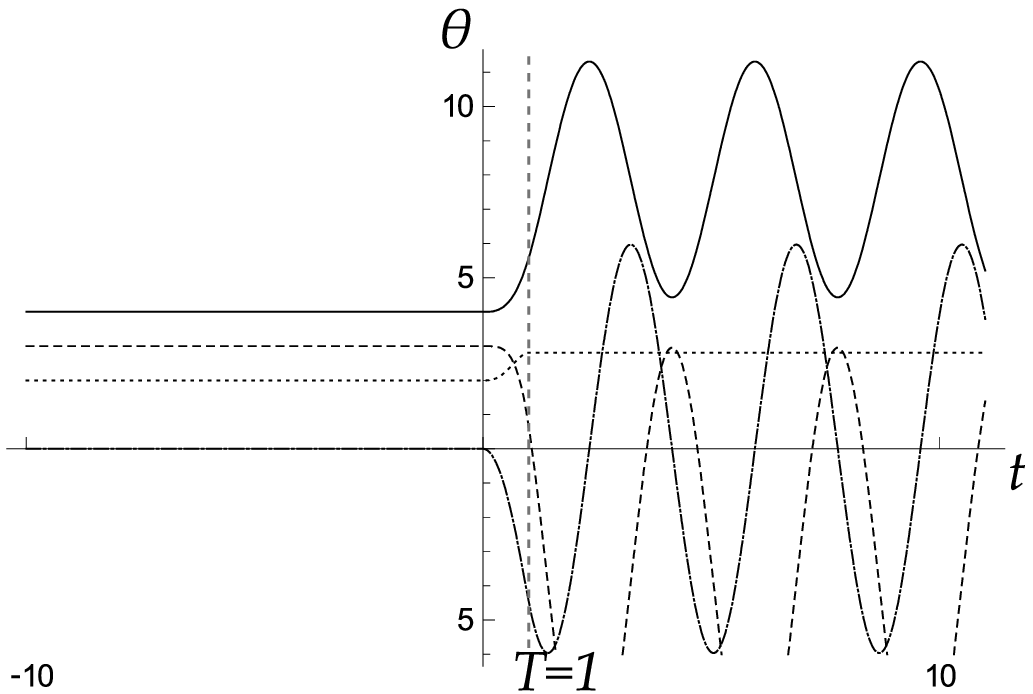}
\includegraphics[width=0.45\textwidth]{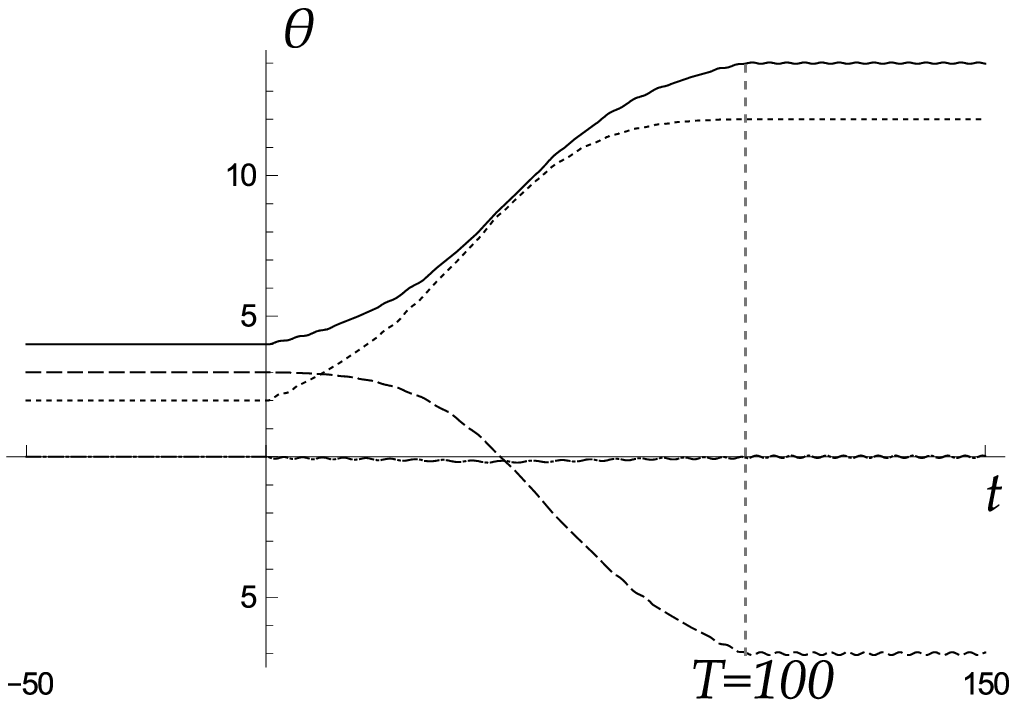}
\end{center}
\caption{Illustration of the evolution of metric of the discussed 2-dimensional system. For concreteness $\vec w=(0,0,3)$ is kept fixed as well as $v_0$ and $v_3$, the remaining components of $\vec v$ are $v_1(t)=v_2(T-t)=2t/T$ between $t=0$ and $t=T$ (the latter marked by the dashed vertical line in the graphs) and remain constant before and after. The graphs show the evolution of $\vt_0$ and three components of $\vec\vt$, starting from the static solution $\vt\stt$. If the change of $H$ is fast, as shown in the left graph where $T=1$, the metric starts to oscillate. On the other hand, an adiabatic change in the left plot ($T=100$ is much greater than the period of the metric of time-independent system) leaves the metric almost in a static state.}
\label{obrasek}
\end{figure*}

\subsection*{Cubic oscillator}

The second model we will ivestigate here is the harmonic oscillator with cubic perturbation on the real line,
\be
H=p^2+q^2+\ii g q^3 \label{iq3h}
\en
with real $g$; $p$ and $q$ are standard position and momentum operators. This Hamiltonian belongs between the most researched non-Hermitian Hamiltonians (except \cite{PTSQM, BBJ} see also older \cite{Cal80}). It is unfortunately not exactly solvable (albeit there is a proof of the spectrum being real \cite{DDT}) but the metric can be calculated perturbatively. It has been done in \cite{BBJ} and we will in principle follow the way outlined there which treats the metric as a function of the operators $p$ and $q$, although for sake of simplicity we will not pursue the most efficient evaluation method that parametrises the metric as an exponential of other operator. Instead we will write the metric directly as
\be
\Theta=\Theta(p,q). \label{thetpq}
\en 
To resolve the problems of operator ordering the parameters $p$ and $q$ in \odk{thetpq} and \odk{iq3h} will be considered $c$-numbers during the calculations, using the Moyal product in place of operator multiplication. (We use the following definition
\be
[F*G](p,q)=F(p,q)\exp{\!\frac{\ii}{2}\!\left[\overleftarrow{\partial_{q}}\overrightarrow{\partial_{p}}-\overleftarrow{\partial_{p}}\overrightarrow{\partial_{q}}\right]}\ G(p,q).\label{moyal}
\en
for the Moyal product; arrows show the direction in which the derivative acts. Its particular advantage is that if an operator $A$  is represented by function $A(p,q)$ then $A^\dag$ is represented by conjugated $A(p,q)^*$.)

Prior to the perturbative evaluation of $\Theta$ we can look at $g=0$ case which is the ordinary harmonic oscillator. Here the time-dependent metrics are obtainable exactly. In chosen representation \odk{tgr} is properly rewritten as
\be
\dot\Theta=\ii(\Theta*H-\ks{H}\ast \Theta)=2(q\frac{\partial\Theta}{\partial p}-p\frac{\partial\Theta}{\partial q}).
\label{rtmp}
\en
It seems favourable to move to polar coordinates in $p$--$q$ plane, so that $p=\rho\sin\vf$ and $q=\rho\cos\vf$. Then the $\rho$-derivatives vanish and \odk{rtmp} becomes
\be
\frac{\partial\Theta}{\partial t}=2\frac{\partial\Theta}{\partial\vf}.\label{lagos}
\en
The general solution to \odk{lagos} can be sought in form
\be
\Theta(\rho,\vf,t)=\Lambda(\rho)\Xi(t+\vf/2).\label{peryjod}
\en
Such function satisfies \odk{lagos} when $\Xi$ is periodic with period $\pi$. Moreover, both $\Lambda$ and $\Xi$ must be analytic if we are to be able to convert them into series of operator products and $\Lambda$ shall be bounded in order to get bounded metric (we can apply stronger criteria for $\Lambda$ to ensure that the metric is well behaved, equation \odk{lagos} puts no restrictions to the $\rho$-dependence of $\Theta$). The static solution here is any function $\Theta\stt$ which depends only on $\rho$; such function corresponds to an operator which commutes with ${H}$ for $H(p,q)=\rho^2$. The periodicity of the harmonic-oscillator metric is clearly result of the equidistance of its spectrum.

For non-zero $g$ one has to rely on perturbative calculation. Due to discussion in \cite{BBJ} a static solution can be obtained by putting
\be
\Theta=1+g \Theta_1(p,q)+g^2 \Theta_2(p,q)+\dots
\en
with $\Theta_i$ being the most general polynomial of order $i+2$. Thus, in the first order in $g$ we get
\be
\Theta_1=\alpha+\beta p+\gamma q+\delta p^2+\ve pq+\zeta q^2+\eta p^3+\vt p^2q+\iota pq^2+\kappa q^3.
\label{poruchansatz}
\en
Substituting it into \odk{tgr} after a rather tedious calculation we get the first order contribution to the static solution, which is
\be
\Ts_1=c+d(p^2+q^2)+pq^2+\frac{2}{3} p^3,
\en
$c$ and $d$ are arbitrary real numbers. Their presence reflects the ambiguity of the metric. In fact, the first two terms correspond to an operator which commutes with $H_0=p^2+q^2$ and hence they are contributions to the free metric $\Theta_0$. Since we wish to start with $\Theta_0=I$ we expect that the adiabatically evolved metric will have $c=d=0$.

Like in case of our 2-dimensional model, we have calculated the time-dependent metric for linear interaction switching, i.e.
\bes
H_\mathrm{I} & = & 0 \quad\mathrm{for}\quad t<0\,,\\
H_\mathrm{I} & = & \frac{\ii tgq^3}{T}\quad\mathrm{for}\quad t\in(0,T)\,,\\
H_\mathrm{I} & = & \ii gq^3 \quad\mathrm{for}\quad t>T.
\ens
For initial condition $\Theta(p,q)=1$ we arrive at
\begin{eqnarray}
\Theta&=&1+\frac{g}{T}\bigg[
\frac{1}{36}(24t-27\sin t+\sin 3t)p^3 \nonumber\\
&&-\frac{4}{3}(2+\cos t)\sin^4\frac{t}{2}p^2q+ \nonumber\\
&&+(t-\frac{3}{4}\sin t-\frac{1}{12}\sin 3t)pq^2 \label{pig}\\
&&-\frac{1}{9}(15+2\cos t+\cos 2t)\sin^2\frac{t}{2}q^3
\bigg]+O(g^2)\nonumber
\end{eqnarray}
for $t\in(0,T)$. Looking at \odk{pig} in $t=T$ one can realise that in the adiabatic limit $T\rightarrow\infty$ only two terms survive; they are equal to $1+g\Ts_1+O(g^2)$ with zero $c$ and $d$ as expected.

\section{Concluding remarks}

We tried to argue that the scattering in \PTm-symmetric quantum theories can be consistently defined by means of an adiabatically switched interaction, and to do it one has to allow time-dependent metrics to be used. Practical calculation of the scattering matrix does not necessarily entail solving \odk{tgr}; it is enough to calculate the static metric for the full Hamiltonian and then proceed with \odk{sfidef}. Some complications may arise from the necessity of calculating the M\o ller operators separately instead of the S-matrix as a whole; consequently one has to make necessary modifications to the standard methods of S-matrix evaluation. More importantly, one must be sure that the metric one has inserted into \odk{sfidef} is the same as the metric which arises from unity via the adiabatic switching. This presents an important restriction since it removes the ambiguity which was often encountered in the \PTm-symmetric theories. (The ambiguity can be removed by demanding simultaneous quasi-Hermiticity of an irreducible set of observables, but in field theory one has usually only the momentum and spin operators which commute with the Hamiltonian and so are useless for the purpose.)

We have argued that the necessary conditions for validity of the described approach are met when the Hamiltonian is diagonalisable and has finite gaps in the spectrum. We do not know to what extent the framework is applicable when continuous spectrum is present. Hopefully there is a consistent generalisation to reasonable class of models with continuous spectrum since there are versions of the adiabathic theorem that do not use the gap condition, see e.g. \cite{avronelgart}.

\bibliography{atdmiptsqt}

\end{document}